\documentclass[
aip, 
cha, 
graphicx,
reprint
]
{revtex4-1}
\usepackage[utf8]{inputenc}
\usepackage[T1]{fontenc}
\usepackage{amsmath}
\usepackage{relsize}
\usepackage[textwidth=15mm]{todonotes}
\usepackage{graphicx}
\usepackage{hyperref}
\usepackage{comment}
\usepackage{ulem}
\usepackage{amsfonts}

\makeatletter
\def\@email#1#2{%
 \endgroup
 \patchcmd{\titleblock@produce}
  {\frontmatter@RRAPformat}
  {\frontmatter@RRAPformat{\produce@RRAP{*#1\href{mailto:#2}{#2}}}\frontmatter@RRAPformat}
  {}{}
}%
\makeatother

\draft 

\providecommand{\norm}[1]{\lVert#1\rVert_{1}}

\newtheorem{theorem}{Theorem}
\newtheorem{definition}{Definition}

\begin{document}


\preprint{AIP/123-QED}



\title{The role of debt valuation factors in systemic risk assessment} 

\author{K. Fortuna}
 \email{kamil.fortuna@pwr.edu.pl}
 \affiliation{Wrocław University of Science and Technology, Faculty of Pure and Applied Mathematics, Hugo Steinhaus Center, Wrocław, 50-370, Poland}
\author{J. Szwabi\'{n}ski}%
 \email{janusz.szwabinski@pwr.edu.pl}
\affiliation{Wrocław University of Science and Technology, Faculty of Pure and Applied Mathematics, Hugo Steinhaus Center, Wrocław, 50-370, Poland
}%

\date{\today}

\begin{abstract}

The fragility of financial systems was starkly demonstrated in early 2023 through a cascade of major bank failures in the United States, including the second, third, and fourth largest collapses in the US history. The highly interdependent financial networks and the associated high systemic risk have been deemed the cause of the crashes. The goal of this paper is to enhance existing systemic risk analysis frameworks by incorporating essential debt valuation factors. Our results demonstrate that these additional elements substantially influence the outcomes of risk assessment. Notably, by modeling the dynamic relationship between interest rates and banks' credibility, our framework can detect potential cascading failures that standard approaches might miss. The proposed risk assessment methodology can help regulatory bodies prevent future failures, while also allowing companies to more accurately predict turmoil periods and strengthen their survivability during such events.
\end{abstract}


\maketitle 

\section{Introduction}

Silvergate Bank, Silicon Valley Bank, Signature Bank. On March 7, 2023, all of them were present on the list of operative banks in the United States; on March 13, there was none. Along with the later collapse of First Republic Bank, the financial system of US experienced the second, third, and fourth largest bank crashes in its history, which occurred in the course of less than two months.
During this crisis, significant uncertainty surrounded the conditions of these institutions. On 12 March, following presidential consultation, three key financial regulatory bodies, the Department of the Treasury, Federal Reserve System, and the Federal Deposit Insurance Corporation, issued a joint statement citing a "systemic risk exception" as the basis for seizing two of these banks.

Systemic risk differs from individual institutional defaults by representing the potential for widespread financial system failure \cite{ilin_uncertainty_2015}. Of particular importance here are critical points, which indicate crucial moments of abrupt qualitative changes, delineating scenarios of smooth system functioning from its complete collapse \cite{jackson_systemic_2020}. The ruin emerges as a  result of
crisis contagion and cascading failures \cite{ilin_uncertainty_2015}, where devaluation of one or a few components leads to the failure of others, 
progressively worsening due to feedback loops created by network dependencies \cite{elliott_financial_2012}.
Understanding and managing systemic risk is crucial for maintaining economic stability and mitigating potential collapse consequences.
Banks worldwide use systemic risk calculations to determine reserve requirements, which in turn influence interest rates and money supply. The fundamental importance of this research area to modern society was recently recognized by the 2022 Nobel Prize in Economics, awarded to Bernanke, Diamond, and Dybvig for their work on bank failures\cite{ diamond_bank_1983}.

\color{black}

Systemic risk models have historically been built on structural financial models, particularly for capturing networks of holdings and transactions between financial institutions. Beginning with Merton's theoretical framework\cite{merton_pricing_1974}, structural asset pricing models use balance sheet data to assess company valuations. These models treat financial instruments issued by companies as derivatives of their balance sheet metrics, applying established methods such as Black-Scholes option pricing formulas to evaluate company stocks\cite{merton_pricing_1974}. 
This approach was naturally extended to networks of cross-holding companies through methods developed by Furfine\cite{furfine_interbank_1999}, Eisenberg-Noe\cite{eisenberg_systemic_2001}, and Bardoscia\cite{bardoscia_debtrank_2015}, as financial instruments held constitute a part of a company balance sheet data. 

While these established algorithms have provided foundational principles for understanding network interactions and crisis propagation, they exhibit limited flexibility due to their simplified treatment of financial instruments. Factors such as recovery rate, time structure of debt, credit quality, and interest rates have been heavily neglected, despite being fundamental components of debt valuation in financial mathematics \cite{bielecki_credit_2004, allen_financial_2013, jarrow_pricing_1995}.
Even though recovery rate has been introduced to several algorithms, it is often ommited in calculations \cite{bardoscia_pathways_2017, barucca_network_2020, european_central_bank_interconnected_2019}. Recovery plays a significant role in models of Eisenberg-Noe\cite{eisenberg_systemic_2001} (EN), Suzuki\cite{suzuki_valuing_2002}, and Fischer\cite{fischer_noarbitrage_2014}, however they focus on the process of clearing, determining creditor claims during defaults, which naturally emphasizes debt retrieval rather than crisis contagion 
 and system collapse
\cite{barucca_network_2020}. 
It is also present in work of Furfine \cite{furfine_interbank_1999} which exhibits similar drawbacks, as valuation dynamics there occurs only in case of actual default occurrence. Nevertheless, recovery is an important component of credit assessment particularly for government-guaranteed banks. Regulators typically intervene to prevent complete loss of citizens' money and financial system collapse, as evidenced by the Swiss Finance Minister's comment regarding the UBS takeover of Credit Suisse: ``I myself am grateful as a customer that this worked out.''\cite{keaten_credit_2023}

The maturity component reflects the flow of time and allows for differentiation across various tenors \cite{bielecki_credit_2004}. This feature is notably absent in previous methods including DebtRank \cite{bardoscia_debtrank_2015, bardoscia_distress_2016, bardoscia_pathways_2017}, with geometric Brownian motion model of external assets dynamics presented in Barucca et al.\cite{barucca_network_2020} constituting a significant exception. 
To provide an illustrative example, mutliplying the remaining time to maturity  by $a$ effectively applies the valuation function multiplier $a$ times, and typical bond maturities can range between three months and ten years, while also approaching one day and more than thirty years \cite{tuckman_fixed_2022}.

There is also a general lack of flexibility to model the diverse credit qualities, which is a fundamental component that allows to valuate bonds of various companies according to their creditworthiness \cite{bielecki_credit_2004, allen_financial_2013, jarrow_pricing_1995}. Fixed valuation rules fail to differentiate between distinct situations: largest bank of the most thriving economy features the same devaluation dynamics as newly created local credit institution operating in a decaying financial system.

Future cashflows are commonly valuated by discount factors determined by the interest rates \cite{bielecki_credit_2004, tuckman_fixed_2022},
which are 
by definition determined by the pricing dynamics of debt in the interbank lending market \cite{tuckman_short-term_2023, klingler_burying_2019}. This in turn shapes the general credit market and further impacts lending policies of banks. These interconnections form a crucial mechanism that drives the dynamics of the financial system and can help explain credit cycles characterized by sudden downturns, as demonstrated in our previous work \cite{fortuna_unified_2023}.

Our research aims to integrate essential debt valuation factors into systemic risk analysis and assess their impact on the US financial system's stability. We focus particularly on identifying critical points where cascading failures might occur due to system interdependencies. We achieve this by incorporating a robust class of reduced-form models\cite{heath_bond_1992,cox_theory_1985,jarrow_pricing_1995,allen_financial_2013, hull_options_2018} into the comprehensive NEVA framework\cite{barucca_network_2020}. In particular, our approach relies on  the following reduced-form model expression for debt valuation:
\begin{equation}
\exp(-[r_t + \lambda_t(1 - \delta_t)](T - t)),  
\label{eq:rf_general}
\end{equation}
where $r_t$, $\lambda_t$, $\delta_t$, and $T$ represent the interest rate, default intensity, recovery rate, and debt maturity, respectively. Based on this formula, we introduce specific parameterized forms of these components into systemic risk calculations and evaluate their implications for financial system stability. We will use the linear DebtRank method\cite{bardoscia_debtrank_2015} as a benchmark, demonstrating how our framework can detect cascading failures that standard approaches might miss.

The proposed modeling approach incorporates the essential features of the debt valuation  in systemic risk calculation on the one hand, and extends the debt pricing process to include a contagion mechanism on the other. It achieves a new level of precision in assessing financial stability, not only improving the robustness of systemic risk calculations, but also extending the applicability of debt valuation models to previously unaccounted situations.

As the construction of our new framework starts from the general reduced-form formula of debt valuation in Eq.~(\ref{eq:rf_general}), it is worth noting that the framework is opened to further extensions, particularly through integration with the existing broad family of models for valuation spreads $ \lambda_t $ and interest rates $ r_t. $  Since the reduced-form models  are the industry standard for pricing fixed-income instruments \cite{tuckman_fixed_2022, bielecki_credit_2004, cox_theory_1985, hull_options_2018}, this approach will facilitate the incorporation of systemic risk calculations into already established business processes.

This article is structured as follows. Section~\ref{sec:Methods} outlines the methods employed in the research, both the existing approaches (Sec.~\ref{subsec:existing_models}) and our contribution (Sec.~\ref{subsec:authors_contribution}). 
In Section~\ref{sec:results}, the security of the American financial system is assessed using the developed models, with particular emphasis on their impact in comparison to existing methods.
Section~\ref{sec:discussion} discusses the results, while Appendix~\ref{subsec:Calibration} provides details of the model calibration.

\section{Methods} 
\label{sec:Methods}

\subsection{Existing methods}
\label{subsec:existing_models}

\subsubsection{Reduced-Form Models}
\label{subsec:rf_models}

Reduced form models constitute a class of models widely used in financial mathematics, particularly in the context of credit risk assessment \cite{allen_financial_2013, hull_options_2018}. These models are designed to capture the dynamics of default events by bypassing explicit modeling of the underlying economic factors that contribute to defaults. Rather than delving into the intricacies of these economic factors, these models concentrate on directly characterizing the timing and incidence of defaults \cite{bielecki_credit_2004, jarrow_pricing_1995, cox_theory_1985}.
Specifically, default events are assumed to follow a Poisson process that results in the  survival probability of the form 
\begin{equation}
\label{eq:poisson}
    Q \left( t, T \right) = \mathbb{E}_t \left[ \exp \left( - \int_t^T \lambda(u) du \right) \right],
\end{equation}
where $\lambda(t)$ is a hazard rate in the Poisson process at time $t.$ 
A default is represented by the jump occurring with probability $\lambda(t) dt$ over period $dt$.

One of the representatives of the family of reduced models is the Duffie-Singleton model \cite{bielecki_credit_2004}. 
It assumes that the amount of a company value that is recovered upon default is proportional to  current market value of its bond $Z(t,T)$ by the rate of $\delta,$ and therefore equals $\delta Z(t,T)$. As a consequence, the price $Z(t,T)$ can be expressed as
\begin{equation}
\label{eq:duffie_singleton_integral}
    Z(t,T) = Z(T,T)  \mathbb{E}_t \left[ \exp \left( - \int_{t}^{T} \left( r(u) + s(u) \right) du \right) \right],
\end{equation}
where $s(t) = \lambda(t)(1 - \delta)$. This equation describes the dynamic evolution of the zero-coupon risky bond price, incorporating the term structure of interest rates $r(t)$ and the credit risk component $s(t)$. 
Under the assumption of time independence of default intensities $s(t) \equiv s$ and instantaneous forward rates $r(t) \equiv r,$ a particularly simplified version of the formula (\ref{eq:duffie_singleton_integral}) is attained:
\begin{equation}
\label{eq:duffie_singleton}
    Z(t,T) = Z(T,T)   \exp \left( - \left( r + s \right) \left( T - t \right) \right).
\end{equation}
This assumption proves advantageous when time $t$ remains fixed and serves predominantly for discounting purposes, while at the same time the dynamic evolution of rates and intensities in time is not an important point of modeling interests. This simplification renders formulas more accessible and practical to manipulate, while preserving essential model aspects simultaneously.
The inclusion of recovery rates in spreads serves to streamline the mathematical formulations, 
making the formulas clearer, more concise, and easier to use, especially compared to the other well-known reduced form model of Jarrow and Turnbull.
\cite{jarrow_pricing_1995}

\subsubsection{NEVA Framework}

Network Valuation in Financial Systems (NEVA), introduced and described in detail in Barucca et al. \cite{barucca_network_2020}, constitute a versatile framework designed for asset valuation within financial networks, analytically separating existing approaches into  asset dynamics and network propagation components.
Many important methods of systemic risk calculation,  including algorithms of Furfine \cite{furfine_interbank_1999}, Eisenberg-Noe \cite{eisenberg_systemic_2001}, and linear DebtRank \cite{bardoscia_correction_2015}, are special cases of NEVA \cite{barucca_network_2020}.
The framework describes balance sheets of $N$ financial institutions and their evolution over time. Balance sheets are divided into external and internal segments.
For a given bank $i$,  $ A_{ij} (t)$ represents its internal assets issued by $j$, while $L_{ij} (t)$ stands for its internal liabilities, possessed by $j$. On the other hand, 
external assets $ A_{i}^{e} (t)$ and liabilities $L_{i}^{e} (t)$ of a bank $i$ constitute a part of its balance sheet that is not included in balance sheets of any other institution of the considered financial system.
At time $t$, banks perform  valuation of their  assets, according to the rules determined by \textit{valuation functions} $\mathbb{V}_{i}^{e} (\cdot)$ and $\mathbb{V}_{ij} (\cdot)$ for external and internal assets, respectively.
In order to properly model debt market dynamics, function space is restricted to \textit{feasible valuation functions}: 
\begin{definition}
\label{def:feasible_valuation}
Function $\mathbb{V} \colon \mathbb{R}^q \to [0, 1]$ is called a feasible valuation function if it is nondecreasing and continuous.
\end{definition}
Then, from the definition of balance sheet components, equity of a bank $i$ $E_i (t)$ can be expressed as
\begin{widetext}
\begin{equation}
\label{eq:neva}
    E_i (t)  = A_{i}^{e} (t) \mathbb{V}_{i}^{e} \left( \mathbf{E} (t)  \right) - L_{i}^{e} + \mathlarger{\sum}_{j=i}^{n} A_{ij} (t) \mathbb{V}_{ij} \left( \mathbf{E} (t) \right) - \mathlarger{\sum}_{j=i}^{n} L_{ij} (t) =:\Phi_i \left( \mathbf{E} (t) \right), \hspace{1cm} i = 1,...,N.
\end{equation}
\end{widetext}

The set of equations (\ref{eq:neva}) can be expressed more concisely as

\begin{equation}
\label{eq:phi_solution}
    \mathbf{E} (t) = \Phi \left( \mathbf{E} (t) \right).
\end{equation}
Consequently, the process of valuation reduces to resolving the fixed-point problem for the map $\Phi.$
Moreover, there exist a correspondence between every solution $\mathbf{E}^* (t)$ of Eq. (\ref{eq:phi_solution}) and a fixed point of the iterative map
\begin{equation}
\label{eq:phi_iterations}
    \mathbf{E}^{(k+1)} (t) = \Phi ( \mathbf{E}^{(k)} (t) ).
\end{equation}
The latter defines the Picard iteration algorithm, providing a method to determine the solutions of the fixed point problem (\ref{eq:phi_solution}). Indeed, $\lim_{k \rightarrow \infty} \mathbf{E}^{k}(t) = \mathbf{E}^{*}(t),$ where $\mathbf{E}^{*}(t)$ is a solution of Eq.~(\ref{eq:phi_solution}) \cite{barucca_network_2020}.

This fact provides a practical approach for conducting stress test analysis, or, in other words, analysing the influence of shocks on the system. Given the initial state of the system, the shock $a$ is applied to external assets, resulting in a dynamics as $A^{e}_i (t) \rightarrow (1-a)A^{e}_i (t).$
Subsequently, employing the Picard iteration algorithm, the propagation of the crisis is computed, enabling assessment of the initial shock's effect.
For any specified precision $\epsilon > 0$, there exists $K(\epsilon)$ such that for all $k > K(\epsilon)$, it holds that $|| \mathbf{E}^{k}(t) - \mathbf{E}^{*}(t) || < \epsilon$. However, it is worth to point out that  that $\epsilon$ is typically unknown beforehand. 
Hence, in practical scenarios, it is reasonable to estimate $\epsilon$ by examining the difference between consecutive terms of the sequence, ensuring it meets the predefined precision criteria, i.e., $|| \mathbf{E}^{k+1}(t) - \mathbf{E}^{k}(t) || < \epsilon$.

The linear DebtRank \cite{bardoscia_debtrank_2015} is a popular method for calculating systemic risk \cite{jackson_systemic_2020}. Although the algorithm predates the establishment of the NEVA framework, it is, in fact, a special case of it \cite{barucca_network_2020}, with the valuation function of the form
\begin{align}
\label{eq:linear_dr}
    \mathbb{V}_{i}^e (E_j (t)) &= 1, \nonumber \\ 
    \mathbb{V}_{ij}(E_j (t)) &= \min \left[ \frac{E_j^+ (t)}{E_j (0)}, 1 \right].   
\end{align}
This method is widely applied in research focused on assessing the stability of financial systems \cite{battiston_debtrank_2012, bardoscia_debtrank_2015}. Notable applications include risk assessments of the Mexican \cite{poledna_multi-layer_2015} and European \cite{european_central_bank_interconnected_2019} banking systems, which involve layers of dependencies such as interbank and external loans, securities cross-holdings, overlapping portfolios, as well as derivatives and foreign exchange dynamics. Additionally, DebtRank has been utilized to tackle risk minimization problems related to common exposures to government bonds \cite{pichler_systemic_2021} and networks of credit default swaps \cite{leduc_systemic_2017}.
Given its significance within the field, the linear DebtRank will serve as a basic reference point throughout this article.
\color{black}

\subsection{Our contribution}
\label{subsec:authors_contribution}

From a modeling perspective, our contribution lies in the incorporation of essential debt valuation components—namely, debt time structure, recovery rate, credit quality, and interest rates—into the analysis of financial networks. This will be achieved by integrating the reduced-form models \eqref{eq:duffie_singleton} into the NEVA framework \eqref{eq:neva}, resulting in a valuation functions of the form:
\label{eq:rf_valuation_function}
\begin{align}
    \mathbb{V}_{i}^e (E_j (t)) &= \exp  \left( - r_{t} \left( T - t \right)  \right), \nonumber \\ 
    \mathbb{V}_{ij}(E_j (t)) &= \exp  \left( - \left(r_{t} + s_t \right) \left( T - t \right)  \right).    
\end{align}

In this approach, distinct models are characterised by the particular definitions of $s_t$ and $r_t$. 
Sec.~\ref{subsec:rf_dr_met} will illustrate the construction of the $s_t$,  a crucial component allowing to connect the credit spread factors with crisis propagation. This model will be enhanced in Sec.~\ref{subsec:ir_feedback_met} by a specified dynamics of interest rates $r_t$, capturing the feedback loop relationships between banking system and general credit market. 
From this method, the linear approximation in Sec~\ref{subsec:rec_dr_met} will allow to derive a simplified model being a generalisation of linear DebtRank at the same time.

\subsubsection{Reduced Form Network Valuation Model}
\label{subsec:rf_dr_met}

Starting with the Duffie-Singleton approach \cite{bielecki_credit_2004} (Eq.~\ref{eq:duffie_singleton_integral}) to recovery, a model 
of risk spread $s_t$ from Equation~(\ref{eq:rf_valuation_function})
will be constructed in this section  in the context of network valuation.
The  spread $s_t$ is represented by $\lambda_t(1 - \delta_t),$ where $\lambda_t$ denotes the default intensity at time $t$, and $\delta_t$ characterizes the recovery rate. 
It is often the case that the recovery rate is presumed to constitute a constant proportion.
However, maintaining a constant recovery rate throughout the entire modeling period may not accurately capture the complexities inherent in real-world contexts.
In fact, it is reasonable to posit that the expected recovery varies in accordance with the dynamic market conditions. During periods of prosperity, a higher expected recovery may be anticipated, given the overall health of a company. Conversely, when a company is facing significant stress, the prospect of recovery from default might be considerably diminished. %
Therefore, it will be modeled as $ \delta_t = \beta \frac{A_j (t)}{A_j (0)}.$
This representation highlights the recovery rate's dependence on the ratio of current assets to initial assets, offering a flexible and realistic depiction of recovery dynamics in our proposed model.
The approach bears resemblance to recovery mechanisms in clearing algorithms \cite{eisenberg_systemic_2001, suzuki_valuing_2002}, where it is defined as the ratio of a company's assets to liabilities in the event of default. 
Thus, it reflects the cautious and recovery-oriented nature of clearing models to some extent, while also being tailored to the requirements of ex ante shock propagation algorithms. 

Within the linear DebtRank model, the default probability is expressed as $ p_j (t) = 1 - \frac{E_j^+ (t)}{E_j (0)}$ \cite{bardoscia_distress_2016}.
When fitting it into the framework of reduced-form models, the default intensity $\lambda_t$ could be depicted analogously as $\lambda(t) = \gamma_j \left(1 - \alpha_j \frac{E_j^+(t)}{E_j(0)} \right)$.
Therefore, the full formula for
the risk spread $s_t$, with minima introduced in order to satisfy the feasibility conditions in Def.~\ref{def:feasible_valuation}, presents as follows:
\begin{widetext}
\begin{equation}
\label{eq:rf_dr_noncal}
    s_t := \max \left[ \gamma_j  \left(1 - \alpha_j \frac{ \min \left[ E_j^+ (t), \ E_j (0)/ \alpha_j \right]}{E_j (0)} \right) \\
     \left(1 - \beta_j \frac{\min \left[ A_j^+ (t), \ A_j (0)/ \beta_j \right]}{A_j (0)} \right), 0 \right].
\end{equation}
\end{widetext}
Here, $\beta_j \in [0, 1]$ is a recovery scale parameter, $\alpha_j \in [0, 1]$ is responsible for the degree of dependence of the hazard rate on the equity, while $\gamma_j \in \mathbb{R}_+ $ describes the overall rate of devaluation.
Assuming no negative equity and assets, $s_t$ satisfies
\begin{equation}
\label{eq:spreads_limits}
    0 \leq s_t  \leq \gamma_j.
\end{equation}
As a result, the minimum value of assets in the model equals $\mathbf{A}(0) e^{- \gamma_j (T-t)}$. 

The parameter $\gamma_j$ governs the rate of decline in asset value and establishes its minimum threshold.
In the absence of the $\gamma_j$ parameter ($\gamma_j \equiv 1$), the model has significantly limited capability to cover scenarios of actual defaults, which is also the case in DebtRank.
The individual values $\gamma_j$ assigned to each company $j$ highlight the diverse credit qualities exhibited by various entities. 
This is a fundamental criterion for pricing debt securities in finance, enabling differentiation between bonds issued by different financial entities, each characterized by distinct risk profiles.

The term $\left(1 - \alpha_j \frac{E_j^+(t)}{E_j(0)} \right)$ captures the inherent dependence of a company's default intensity on its equity value. This relationship stems from the definition of default as the point in time $\tau$ when a company's assets $A_i(\tau)$ fail to cover its liabilities $L_i(\tau)$, leading to $E_i(\tau) := A_i(\tau) - L_i(\tau) \leq 0$. Consequently, the connection between the probability of a company's default and the current value of its equity is straightforward. 
It constitutes a fundamental aspect of systemic risk research. 
An initial decline in a company's equity value results in an elevated probability of default, precipitating the devaluation of its issued financial instruments. Consequently, holders of these instruments experience a decrease in asset value, equity, creditworthiness, and the value of their securities. This sets off a feedback loop, amplifying the crisis across interconnected entities within the market network. The iterative propagation mechanism is indispensable to the systemic risk discipline, distinguishing it from conventional financial mathematics, where market dependencies are typically represented as one-time correlation effects or shared exposures to a common factor like the market portfolio \cite{markowitz_portfolio_1952, vasicek_finance_2015}.
In the model developed herein, the feedback loop is also ingrained in the dynamics of the recovery rate through its reliance on asset value. This integration not only facilitates the inclusion of debt recollection mechanisms in systemic risk calculations, but also enables it to reflect the behavior characteristic for crisis phenomena.

After the process described in Appendix~\ref{subsec:Calibration}, the spread dynamics is characterised by

\begin{widetext}
\begin{equation}
\label{eq:rf_dr}
    s_t = \max \left[ \gamma_j  \left(1 -  \frac{ \min \left[ E_j^+ (t), \ E_j (0) \right]}{E_j (0)} \right) \\
     \left(1 - \beta_j \frac{\min \left[ A_j^+ (t), \ A_j (0)/ \beta_j \right]}{A_j (0)} \right), 0 \right].
\end{equation}
\end{widetext}

\subsubsection{Interest Rates Feedback Model}
\label{subsec:ir_feedback_met}

In order to incorporate interest rates to the dynamics of financial networks, it is assumed that they are inherently integrated into the valuation from the outset and are factored into the calculation of initial assets $\mathbf{A}(0)$. Consequently, $r_{t} = r_0 + \Delta r_t$ is introduced, which leads to 
\begin{align}
    \mathbb{V}_{i}^e (E_j (t)) &= \exp  \left( -\Delta r_{t} \left( T - t \right)  \right), \nonumber \\ 
    \mathbb{V}_{ij}(E_j (t)) &= \exp  \left( - \left(\Delta r_{t} + s_t \right) \left( T - t \right)  \right).
\label{eq:rf_valuation_functions}
\end{align}
In the context of financial modeling, the simplistic approach of treating interest rates  as exogenous variables may fall short in accurately representing the intricacies of real-world dynamics. Extensive empirical evidence underscores the existence of complex relationships between the banking system and interest rates \cite{ huser_how_2021, bowman_how_2020}. Detailed discussions regarding the motivations and rationale behind those connections are elaborated in~Sec~\ref{subsec:IR_feedback_results}.
Henceforth, within the framework of our model, the interest rate may depend on the condition of the banking system, as represented by the value of the equity vector. This relationship is formally expressed by:
\begin{equation}
\label{eq:ir_general_equties}
    \Delta r_t = \Delta r_t \left( \mathbf{E} (t) \right).
\end{equation}

Up to this point, the valuation function $\mathbb{V}_{ij}$ in various models has been tied solely to the equities $E_j (t)$ of counterparties $j$ undergoing credit evaluation. However, the NEVA framework takes a broader approach through the formulation of the map $\Phi$, as defined in Eq. (\ref{eq:neva}). This definition facilitates a more general spectrum of dependencies of valuation functions $\mathbb{V}_{ij}$ on the entire vector of equities $\mathbf{E} (t)$. Consequently, all outcomes within the framework remain applicable in this extended context. Thus, under the modeling assumption presented in Equation (\ref{eq:ir_general_equties}), the framework's results hold true, provided that valuation functions remain feasible.

In order to elucidate the impact of banking system credibility on interest rates, the proposed dynamics is of the form
\begin{equation}
\label{eq:ir_dynamics}
     \Delta r_{t}   =  \widetilde{\gamma} \left(1 - \widetilde{\alpha} \frac{ \norm{ \mathbf{E}^+ (t) }}{ \norm{ \mathbf{E}(0) }} \right) \left(1 - \widetilde{\beta} \frac{ \norm{ \mathbf{A}^+ (t) }}{ \norm{ \mathbf{A}(0) }} \right).
\end{equation}
Here, $\mathbf{u} = \left( u_1, u_2,..., u_n \right) \in \mathbb{R}^n$ and $\norm{\mathbf{u}} = \mathlarger{\sum}_{k=1}^{n} |u_k|$. This proposition is motivated by the fact that fundamental basic interest rates reflect the overall borrowing costs within the interbank lending market \cite{tuckman_short-term_2023, klingler_burying_2019}.
Eventually, the complete characterisation of the reduced form model can by written as
\begin{widetext} 
\begin{align}
\label{eq:ir_feedback}   
    s_t &=  \max \left[ \gamma_j  \left(1 -  \alpha_j \frac{ \min \left[ E_j^+ (t), \ E_j (0) \right]}{E_j (0)} \right)
     \left(1 - \beta_j \frac{\min \left[ A_j^+ (t), \ A_j (0)/ \beta_j \right]}{A_j (0)} \right), 0 \right] \nonumber \\
    \Delta r_t &=  \max  \left[   \widetilde{\gamma} \left(1 - 
    \widetilde{\alpha} \frac{ \min \left\{ \norm{ \mathbf{E}^+ (t) }, \norm{ \mathbf{E}(0) }  \right\} }
    { \norm{ \mathbf{E}(0) }} \right) \left(1 - \widetilde{\beta} \frac{ \min \left\{ \norm{ \mathbf{A}^+ (t) }, \widetilde{\beta}^{-1} \norm{ \mathbf{A}(0) }  \right\}}{ \norm{ \mathbf{A}(0) }} \right)  , 0 \right].
\end{align}
\end{widetext}

Following the process outlined in Appendix~\ref{subsec:Calibration}, the formula takes the form below:
\begin{widetext} 
\begin{align}
\label{eq:ir_feedback_noncal}   
    s_t &=  \max \left[ \gamma_j  \left(1 -  \frac{ \min \left[ E_j^+ (t), \ E_j (0)/ \right]}{E_j (0)} \right)
     \left(1 - \beta_j \frac{\min \left[ A_j^+ (t), \ A_j (0)/ \beta_j \right]}{A_j (0)} \right), 0 \right] \nonumber \\
    \Delta r_t &=  \max  \left[   \widetilde{\gamma} \left(1 - 
    \frac{ \min \left\{ \norm{ \mathbf{E}^+ (t) },  \norm{ \mathbf{E}(0) }  \right\} }
    { \norm{ \mathbf{E}(0) }} \right) \left(1 - \widetilde{\beta} \frac{ \min \left\{ \norm{ \mathbf{A}^+ (t) }, \widetilde{\beta}^{-1} \norm{ \mathbf{A}(0) }  \right\}}{ \norm{ \mathbf{A}(0) }} \right)  , 0 \right].
\end{align}
\end{widetext}

\subsubsection{Recovery DebtRank}
\label{subsec:rec_dr_met}

In this section, a simplified model will be derived from Eq.~(\ref{eq:rf_dr}), focusing  specifically  on the recovery rate impact. The result will effectively be a model that incorporates a dynamic recovery mechanism into linear DebtRank, demonstrating that the latter can be approximated by our model with reduced parameters. 

As the emphasis of the model developed in this section is the recovery rate specifically, interest rates do not constitute an object of interest, and therefore will be disregarded from further consideration.
Setting $ \gamma_j = \left( T - t \right) =1 $ in Eq.~(\ref{eq:rf_dr}) and employing the approximation $e^{-x} \approx 1 - x$ yields
\begin{widetext}
\begin{align}
    \exp \left( - \left(1 - \alpha_j \frac{E_j^+ (t)}{E_j (0)} \right) \left(1 - \beta_j \frac{A_j^+ (t)}{A_j (0)} \right)  \right)  \approx 1 - \left(1 - \alpha_j \frac{E_j^+ (t)}{E_j (0)} \right) \left(1 - \beta_j \frac{A_j^+ (t)}{A_j (0)} \right) 
    \approx  \alpha_j \frac{E_j^+ (t)}{E_j (0)} + \beta_j \frac{A_j^+ (t)}{A_j (0)}.
\end{align} 
\end{widetext}
It leads to 
\begin{equation}
\label{eq:rec_dr_noncal}
    \mathbb{V}_{ij}(E_j (t)) = \left( \alpha_j \frac{E_j^+ (t)}{E_j (0)} + \beta_j \frac{A_j^+ (t)}{A_j (0)} \right).
\end{equation}
After the calibration process described in Appendix~\ref{subsec:Calibration}, one gets
\begin{equation}
    \mathbb{V}_{ij}(E_j (t)) = \left( \alpha_j \frac{E_j^+ (t)}{E_j (0)} + (1-\alpha_j) \frac{A_j^+ (t)}{A_j (0)} \right).
\end{equation}
This effectively establishes a convex combination of two potential valuation rules. 
The first one is based on the probability of default, estimated by the proportion of the current equity relative to the initial one. The second one describes the recovery in case of default and utilizes the ratio of assets. 
Within this framework, the inclusion of a weight parameter $\alpha_j$ allows for the adjustment of the relative importance assigned to the default probability and recovery components. This flexibility enables the model to be tailored to various situations and application-specific needs.
It is worth to point out that the Linear DebtRank emerges as a specific case within this approach when $\alpha_j$ is set to 1. 
Another particularly simple parameterisation applies the same weights of importance to the default probability and recovery components, i.e. $\alpha_j = \frac{1}{2}$, yielding
\begin{equation}
     \mathbb{V}_{ij}(E_j (t)) = \frac{1}{2} \left(  \frac{E_j^+ (t)}{E_j (0)} + \frac{A_j^+ (t)}{A_j (0)} \right).
\end{equation}
This approach offers a direct compromise between the significance of model components, serving as a natural and effective starting point when there is a requirement to incorporate recovery rates into systemic risk calculations, traditionally conducted by the means of the linear DebtRank algorithm~(\ref{eq:linear_dr}).

The final formula presents as follows:

\begin{align}
\label{eq:recovery_dr}
        \mathbb{V}_{i}^e (E_j (t)) &= 1, \nonumber \\
        \mathbb{V}_{ij}(E_j (t)) &= \min \left[ \alpha_j \frac{E_j^+ (t)}{E_j (0)} + (1 - \alpha_j) \frac{A_j^+ (t)}{A_j (0)}, 1 \right].
\end{align}

\section{Results}
\label{sec:results}

\subsection{Dataset}

The proposed modeling framework will be employed to assess the stability of the financial system in the United States. 
The dataset encompasses the top 15 American banks in terms of assets, with the exclusion of Charles Schwab due to data availability limitation. Collectively, the subset represent approximately 58\% of the domestic banking system's assets
and is assumed sufficiently representative of the American banking landscape for the intended research purposes. 
Additionally, the dataset includes information on five prominent mutual funds. The data is sourced from the Eikon Reuters analytical platform, 
providing comprehensive balance sheet details for the analyzed entities, including asset and liability values, as well as information on debt cross-holdings.

\begin{figure}
    \centering
    \includegraphics[width=0.49\textwidth]{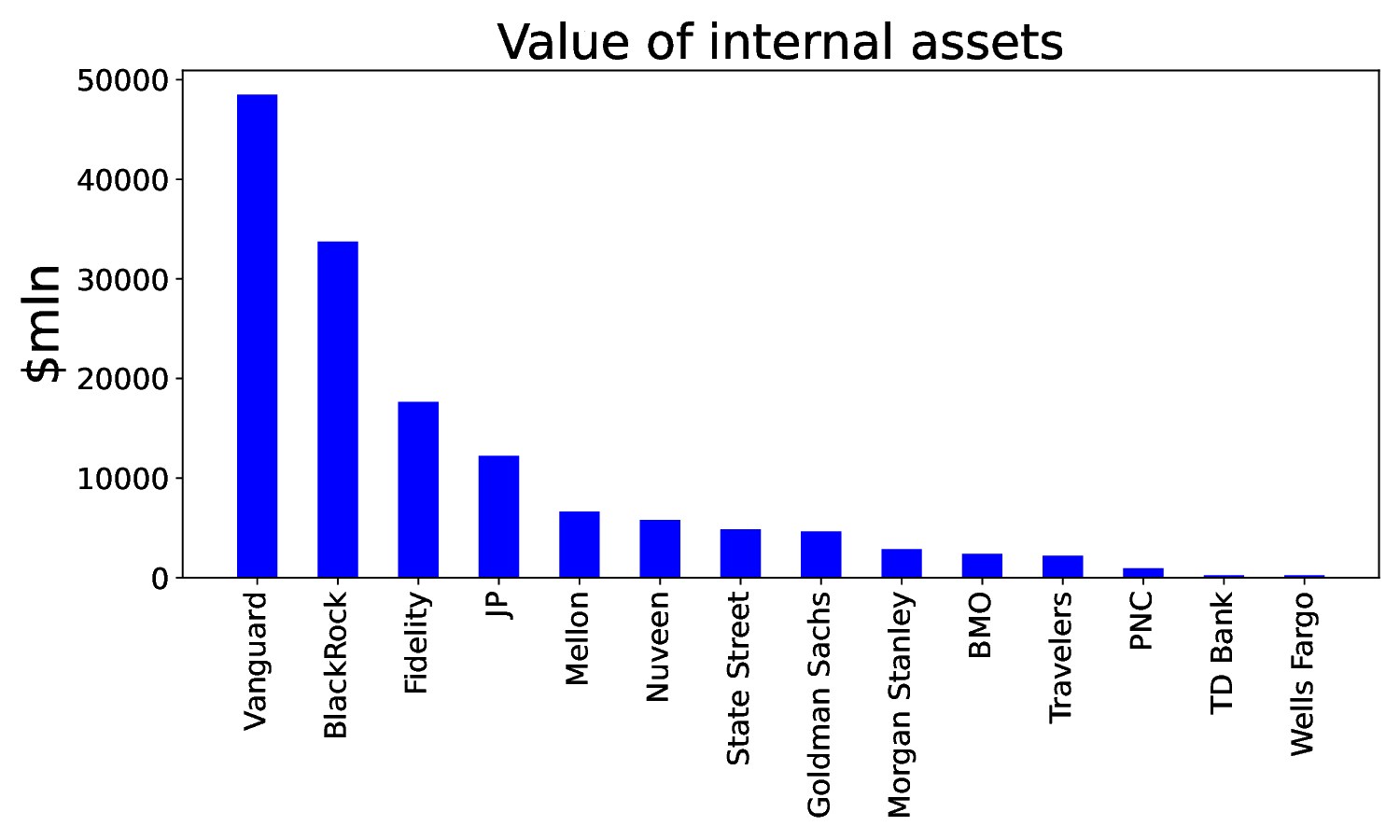} \\\includegraphics[width=0.49\textwidth]{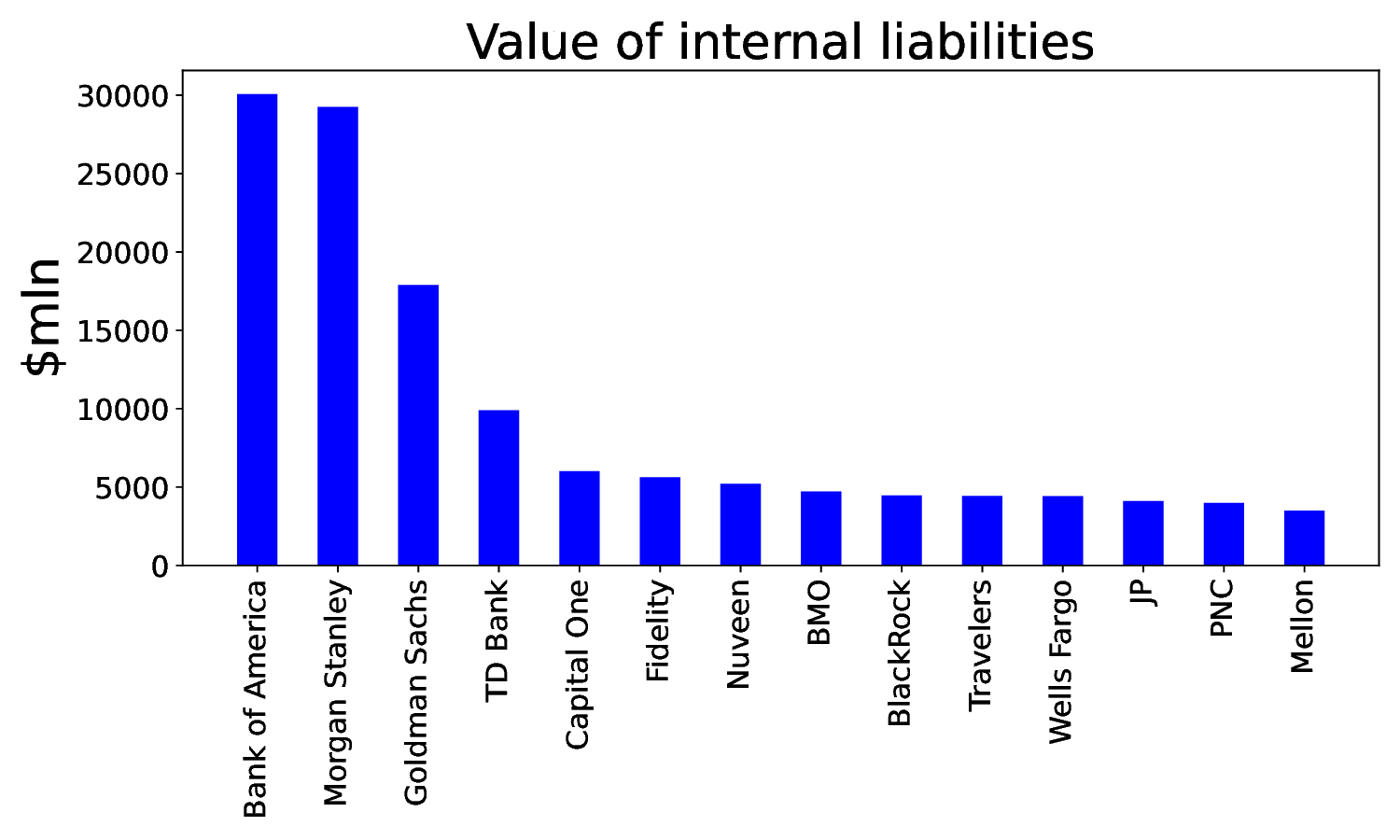} 
    \caption{ Breakdown of internal assets and internal liabilities among top banks and mutual funds in the United States. The top panel demonstrates a descending order of internal asset values, portraying a distribution akin to a typical power law curve. Mutual funds, particularly Vanguard and BlackRock, emerge as prominent stakeholders, holding a significant lead. 
    The bottom panel showcases the hierarchy of internal liabilities among specific companies, revealing notably higher internal debt for the top four cases, gradually diminishing thereafter until the final two companies, which exhibit visibly lower internal indebtedness compared to their counterparts.}
    \label{fig:int_holdings}
\end{figure}

The internal assets and liabilities held by financial entities within the United States are illustrated in Fig. \ref{fig:int_holdings}.
Bank of America and Morgan Stanley emerge as the predominant systemic borrowers, with Goldman Sachs and TD Bank following behind, collectively representing 61\% of the total internal debt within the system. Consequently, they pose significant direct sources of systemic risk within the examined market. Conversely, Vanguard and Citigroup exhibit substantially lower absolute borrowing figures. 
Mutual funds play a significant role among the holders of internal assets, collectively accounting for 75\% of the system's total holdings. Among these, the three largest holders, Vanguard, BlackRock, and Fidelity, hold 70\% of the whole system internal assets. This indicates that within the sample, mutual funds, especially the aforementioned three, are disproportionately exposed to systemic risk as measured by their internal asset holdings. 
As their portfolios decline sharply with worsening condition of the system, their devaluations will exacerbate each other, emphasizing the need for close monitoring. Excluding mutual funds from banking system analyses could lead to biases and overconfidence in system security, resulting in excessive risk-taking and potentially catastrophic outcomes. This mirrors the 2007 crisis, where failure to recognize the housing market's risks and the financial system's exposure, compounded by firms like Lehman Brothers and AIG, led to the worst global financial crisis since the Great Depression \cite{williams_uncontrolled_2010, rosenberg_concise_2012}.
\color{black}

\subsection{Model behavior and its impact on financial system analysis}

\subsubsection{Recovery DebtRank}
\label{subses:rec_dr_res}

In this section, relevance of the recovery rate will be analyzed. Our recovery DebtRank model, introduced in Eq.~(\ref{eq:recovery_dr}),
\begin{equation}
\label{eq:rr_dr_res}
    \mathbb{V}_{ij}(E_j (t)) = 
    \left( \alpha_j \frac{E_j^+ (t)}{E_j (0)} + (1-\alpha_j) \frac{A_j^+ (t)}{A_j (0)} \right),
\end{equation}
will be used for that purpose, as it isolates the recovery rate $\rho_j = (1-\alpha_j) \frac{A_j^+ (t)}{A_j (0)}$ from other effects considered in this work. The results will be compared with the linear DebtRank due to its widespread adoption in the systemic risk community. 
As the general system dynamics is the primary focus of this study, $\alpha_j$ is assumed to be equal across all companies.

\begin{figure}[ht]
\centering
\includegraphics[height=0.28
\textwidth]{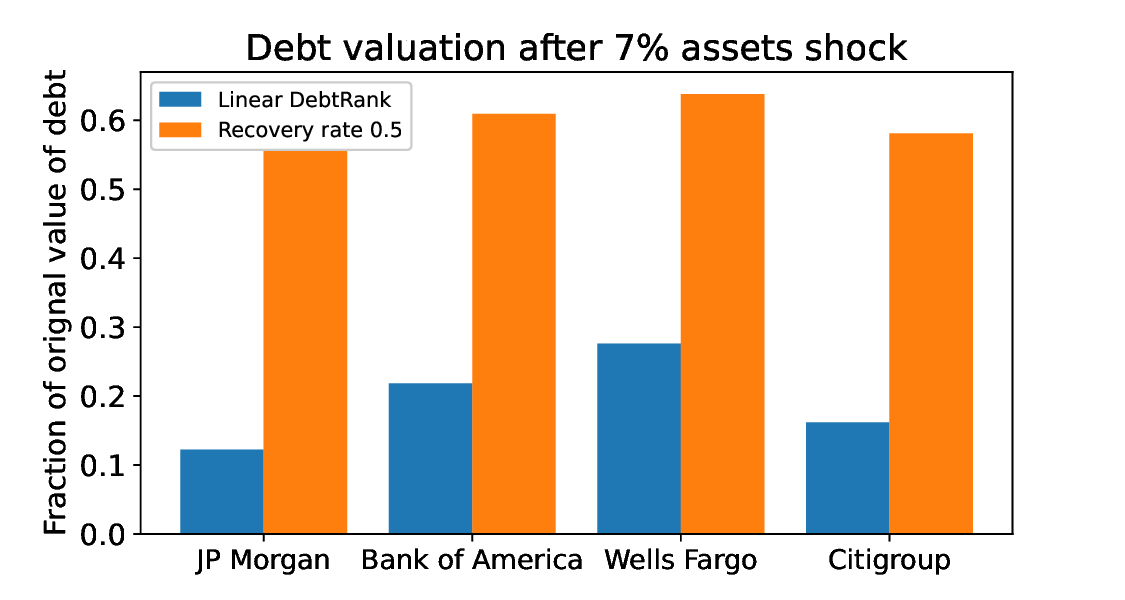} \\
\includegraphics[width=0.49\textwidth]{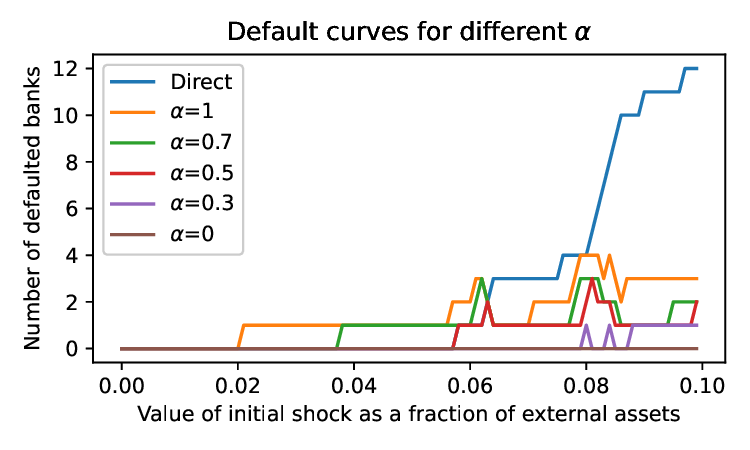}
\caption{The impact of recovery rate on financial system dynamics.
The top panel illustrates the valuation of debt instruments for leading American banks under a 7\% asset shock, as determined by the DebtRank algorithm, considering a recovery rate with $\alpha_j=0.5$. Calculations were conducted for the 'Big Four' banks, a prominent group in the American financial system, collectively representing 47\% of the US bank market.
The bottom panel depicts the relationship between the number of defaulted institutions and varying percentages of shock to external assets, calculated using a recovery DebtRank model (\ref{eq:recovery_dr}). The blue curve represents direct defaults triggered by the shock, while distinct curves illustrate indirect defaults, each corresponding to a different value of the $\alpha_j$ parameter. Since the recovery parameter equals $1-\alpha_j$, the number of defaults increases with $\alpha_j$. For the sake of simplicity, the $j$ subscript has been omitted in the plots since the same $\alpha_j$ was assumed for all companies in calculations.}
\label{fig:RR}
\end{figure}

The influence of the recovery rate on the systemic risk assessment is illustrated in Figure~\ref{fig:RR}.
The impact of varying recovery rates on the outcomes is clearly evident. In the case of linear DebtRank (with no recovery,  $\alpha_j$=1) and lower recovery rate values, a noticeable shift in the series of defaults occurs near the 6\% shock threshold, indicating that systemic risk effects have the potential to alter the boundary of a precarious regime. Around the critical point of 8\%, where the number of direct defaults experiences a sharp increase, lower recovery rates result in a discernible network contagion effect, exacerbating the severity of the crisis. 
In contrast, when recovery rates are higher, the shifts of critical points and crisis amplification are notably less pronounced or virtually absent. 
Moreover, the price dynamics of debt, determined by valuation function in Eq.(\ref{eq:rr_dr_res}) with the absence of recovery ($\alpha_j=1$), may exhibit an exaggeratedly pessimistic trend, as evidenced by the illustration provided at the bottom panel of Fig.\ref{fig:RR}. It elucidates the evaluation of 1\$ of debt from the "Big Four" banks, which collectively constitute 47\% of the American banking market, when their assets are valued at 93\% of their original worth, both with and without the prospect of recovery. The disparities between these scenarios are significant, with the ratio reaching 1:5. In the absence of recovery, precipitous decline to approximately 20\% of original value appears implausible, whereas the results of calculations with $\alpha_j=0.5$ appear far more reasonable, fluctuating around 60\%.
Furthermore, at the brink of default, assets equal liabilities, allowing for full debt repayment. Additionally, deposits are backed by state insurers like FDIC,
and therefore rarely fall completely \cite{resseguie_banks_2020}.
Therefore, the dynamics stemming from linear DebtRank might be inherently artificial, potentially leading to inaccuracies in calculation results. Such inaccuracies, in turn, can distort risk perceptions and compromise the quality of decision-making processes. 
While failing to recognize the danger of a crisis can lead to catastrophic consequences, overly stringent regulation can also yield serious negative effects, particularly due to its prolonged influence and the accumulation of its consequences over time. Such regulations can impede individuals and businesses from accessing credit and financing, as banks may adopt more cautious lending practices, resulting in diminished availability of loans, especially for small businesses and individuals with imperfect credit histories \cite{mcleay_money_2014}. 
Consequently, this can hamper economic growth, hinder job creation, reduce productivity, and dampen overall prosperity \cite{angelina_effects_2020}. Moreover, strict regulations may exacerbate poverty levels, which in turn profoundly impacts people's health by constraining access to healthcare, nutritious food, safe drinking water, and adequate living conditions \cite{poirier_systematic_2024, de_oliveira_incorporating_2022}.

\subsubsection{Reduced Form: Recovery and Creditworthiness.}
\label{subsec:rf_gammabeta}
In this section,  the framework of reduced form models in network valuation will be applied to assess systemic risk within the U.S. financial landscape.
To recall, the dynamics of the risk spread $s_t$ is given by Eq.~(\ref{eq:rf_dr}):
\begin{widetext}
\begin{equation}
\label{eq:rf_dr_res}
    s_t = \max \left[ \gamma_j  \left(1 -  \frac{ \min \left[ E_j^+ (t), \ E_j (0) \right]}{E_j (0)} \right) \\
     \left(1 - \beta_j \frac{\min \left[ A_j^+ (t), \ A_j (0)/ \beta_j \right]}{A_j (0)} \right), 0 \right].
\end{equation}
\end{widetext}

The goal is to evaluate the impacts of recovery parameter $\beta_j$ and credit quality parameter $\gamma_j$. 
Because in the absence of interest component $r_{t}$ in the reduced form model~(\ref{eq:rf_valuation_function}) the influence of maturity $(T-t)$ is qualitatively equivalent to $\gamma_j$, analysis is performed only for $\gamma_j$ parameter.
Moreover,  parameters $\beta_j, \ \gamma_j$ are assumed to be the same across all companies for the sake of simplicity.

\begin{figure}
    \centering
    \includegraphics[width=0.49\textwidth]{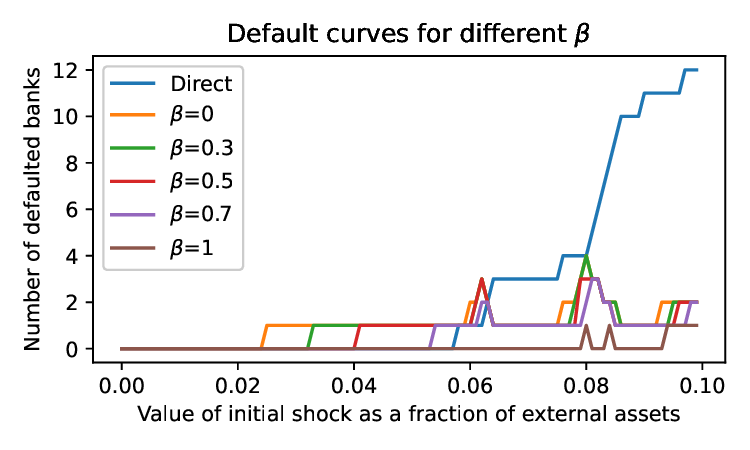} \\
    \includegraphics[width=0.49\textwidth]{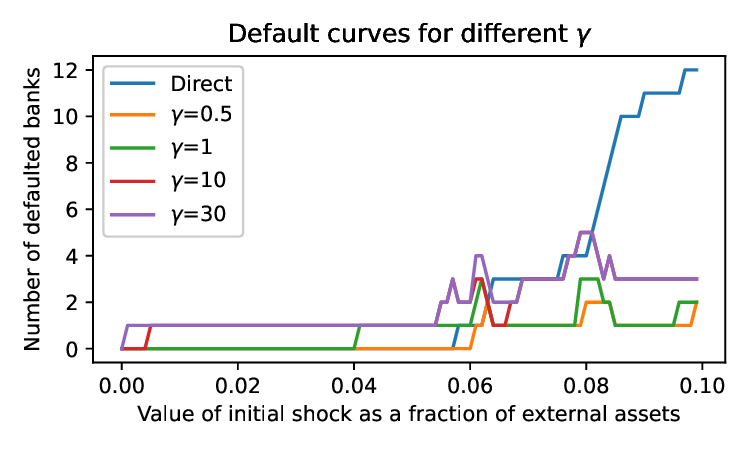}
    \caption{The relationship between the number of defaults and varying initial shocks to external assets.
    The chart illustrates the dependency between the number of defaulted institutions and different percentages of shock to external assets, computed using a reduced-form model (Eq.~ \ref{eq:rf_dr}). The blue curve represents direct defaults triggered by the shock, while distinct curves depict indirect defaults, each corresponding to different parameter values. For the top panel, computations were performed with parameter values $\Delta r_t = 0, \ (T-t) = 1, \ \gamma_j=1,$ and various curves correspond to different values of the recovery parameter $\beta_j$. The quantity of defaults decreases with $\beta_j$ as it determines the assets retrieved in case of default. Similarly, for the bottom panel, computations were conducted with parameter values $\Delta r_t = 0, \ (T-t) = 1, \ \beta_j=0.5,$ and various curves correspond to different values of the scale parameter $\gamma_j$. The amount of defaults increases with $\gamma_j$ as it describes the pace of valuation decline. For the sake of simplicity, the $j$ subscript has been omitted in the plots since the same parameter values were assumed for all companies.}
    \label{fig:rf_params}
\end{figure}

The results are presented in Figure~\ref{fig:rf_params}. It illustrates the relationship between the number of failed companies and the magnitude of external assets' shocks for different values of the model parameters.
The influence of  $\gamma_j$ outweighs the impact of the recovery rate. In order to comprehensively assess the recovery rate, parameter values spanning the entire spectrum, including extreme values of 0 and 1, were examined. 
Given the fact that the number of defaults increase along with the $\beta_j$ parameter, this choice offers a broad understanding of outcomes' dependence on recovery. 
The overall behavior observed in the analysed case is similar to that of the recovery DebtRank (Fig.~\ref{fig:RR}).
Conversely, the dynamics of $\gamma_j$ exhibits a more substantial impact. In the case of recovery, systemic defaults primarily manifest as slight shifts around critical thresholds of 6\% and 8\%, resulting in immediate direct defaults. A minor amplification of the crisis, evidenced by two predominant indirect defaults, occurs within the range of shock magnitudes between 9\% and 10\%. This amplification occurs toward the end of the analyzed shock values, when the number of direct defaults reaches the point of 10, while up to that moment in the simulation there is only one prevailing indirect default. The effect magnitude is stronger in case of recovery DebtRank, with two prevailing indirect defaults from the point of 7\% and three ones after the critical point of 8\%, as depicted in Fig.~\ref{fig:RR}.
Nevertheless, the impact stemming from variations in $\gamma_j$ is more pronounced. Effects occurring around critical thresholds of 6\% and 8\% exhibit heightened severity, evidenced by a larger number of defaults and more substantial shifts. Moreover, following the critical threshold of 6\%, the amount of indirect defaults remains consistently around 3, significantly contributing to the ongoing amplification of the crisis. In fact, from the very beginning of crisis around the 6\% point, the network effect at least doubles the crisis magnitude most of the time up to the 8\% threshold of sharp increase in direct defaults.
This underscores the significance of the $\gamma_j$ parameter and, by extension, the incorporation of reduced-form models into the network valuation framework. 

\subsubsection{System boundaries}
\begin{figure}[ht]
\centering
\includegraphics[width=0.49\textwidth]{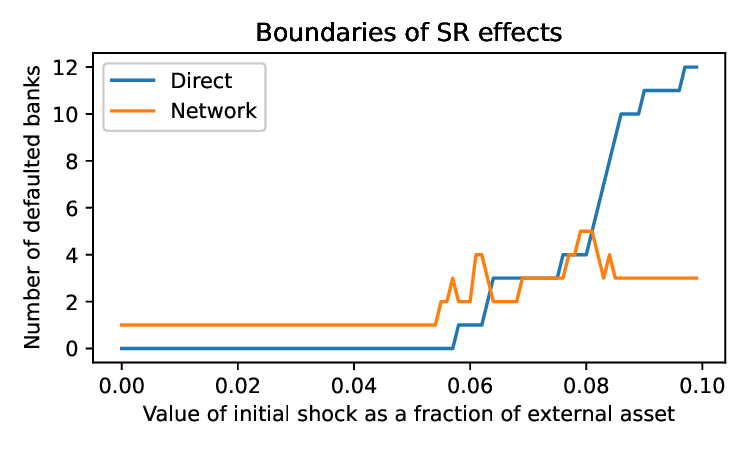}
\caption{The boundaries of contagion effect.
The figure illustrates the number of direct and boundary of indirect defaults for various initial shocks to external assets with static interest rates ($\Delta r_t \equiv 0$). Two prominent spikes in the default series are evident, occurring around shock magnitudes of 6\% and 8\%. At the 6\% shock magnitude, systemic risk influence becomes apparent, with peaks indicating three systemic defaults in the case of zero direct failures and four in the case of one, respectively. The network effect diminishes as direct defaults surge sharply around the 6\% threshold but regains momentum approaching the 8\% shock magnitude, coinciding with another peak wherein four direct and five indirect defaults are observed. Subsequently, the number of systemic defaults begins to decline, stabilizing at three, while direct defaults progressively dominate, peaking at 12 at the 10\% shock magnitude.}
\label{fig:sr_boundaries}
\end{figure}

The boundaries of the number of defaults in the analysed network topology are depicted in  Fig.~\ref{fig:sr_boundaries}. Notably, these boundaries closely align with the curve corresponding to the highest examined parameter value $\gamma_j = 30$ of the reduced form model, as demonstrated in Figure~\ref{fig:rf_params}.
In the context of topology of financial system under consideration, network effects primarily induce shifts in crises and amplify pre-existing ones. 
A noticeable shift occurs around the point of 6\% shock magnitude, important to consider when assessing system stability.
However, this shift does not substantially alter system behavior. Based on these findings, there is insufficient evidence to posit that systemic effects serve as a primary source of severe crises; rather, they amplify or increase probability of occurrence for crises caused by direct shocks, without fundamentally altering the system's nature. Consequently, there is a basis to infer that network effects do not inherently pose a qualitative danger to the financial system of United States. It seems reasonable to state that, for instance, the incorporation of safety margins into critical point calculations conducted by standard non-network techniques is able to provide a sufficiently accurate assessment of secure area boundaries, without the actual need to resort to systemic risk methods. Nevertheless, financial systems are complex systems, with multiple layers of intertwining relationships.

\subsubsection{Interest rates feedback loop and cascading failure}
\label{subsec:IR_feedback_results}

The onset of the coronavirus pandemic prompted unprecedented government interventions worldwide, with substantial aid packages injected into economies at an unprecedented scale. Notably, the CARES Act emerged as the largest stimulus measure in U.S. history, amounting to 10\% of the country's GDP. However, the influx of funds and pandemic-related restrictions was followed by an inflationary surge.
The primary mandate of central banks, notably the Federal Reserve System, includes preserving financial stability, particularly the value of the national currency. Consequently, the Federal Reserve was compelled to raise interest rates in order to curb escalating inflation. 
Although banks maintained fairly conservative portfolios, characterized by significant holdings in mortgage-backed securities and U.S. Treasuries, traditionally considered reliable until maturity, they were nevertheless highly vulnerable to fluctuations in interest rates. Such fluctuations profoundly impact the valuation of future cash flows associated with these assets. Consequently, an increase in interest rates precipitated a sharp decline in the market values of these instruments, thereby eroding the capital reserves of banks holding such assets. To sustain liquidity, some banks were compelled to sell their holdings, resulting in significant losses.
Over the course of a few days, between 8 and 12 March, Silvergate Bank, Silicon Valley Bank, and Signature Bank collapsed, which, along with the subsequent failure of First Republic Bank on 1 May, represented the second, third, and fourth largest bank failures in the history of the United States. 
This sequence of events has resulted in a direct impact on spreads within the financial ecosystem. 

\begin{figure}[ht]
\label{fig:sofr}
\centering
\includegraphics[width=0.49\textwidth]{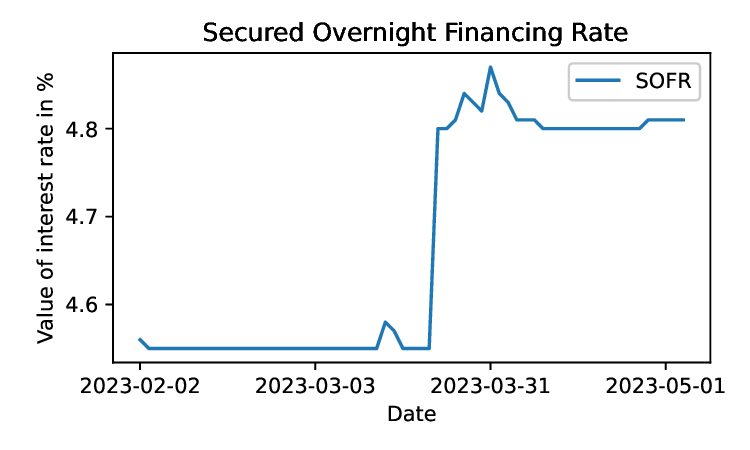}
\caption{The  evolution of the Secured Overnight Financing Rate (SOFR) over time. Vertical lines represent shifts in Federal Reserve policies regarding federal funds rates. Under a fixed FED policy, the interest rate value remains relatively constant. Two distinctive areas are particularly noticeable: a 3bps relative spike on March 15th and a period of intensive oscillation between March 24th and April 5th, with a peak of 7bps relative to the neighboring plateau observed on March 31st. Source: \url{https://fred.stlouisfed.org/series/SOFR}.}
\end{figure}

This effect 
is visible in Fig.~\ref{fig:sofr}, which illustrates the evolution of The Secured Overnight Financing Rate (SOFR)
from February to May 2023. 
SOFR represents the expense of overnight cash borrowing, collateralized by U.S. government bonds, and is derived from transaction data of the repurchase agreements market \cite{klingler_burying_2019}. 
It replaced LIBOR as a benchmark rate, therefore constituting a crucial component of the American (and therefore global) financial system, serving as a foundational metric for valuing bonds, derivatives, mortgages, and loans \cite{tuckman_fixed_2022}.
Following the described series of banks collapses, the value of SOFR spiked by 3bps to 4.58\% on 15 March, compared to the otherwise mainly stable level of 4.55\%. This surge reflects heightened market uncertainty stemming from financial turmoil, prompting banks to demand higher credit prices due to increased risk.
As the United States remains a key node in the global financial system, built on the foundations of Bretton Woods Agreement, disturbances spread across the globe, impacting foreign institutions that issue instruments traded on American stock markets and held by American financial companies.
On 19 March, Credit Suisse, the second largest bank in Switzerland, underwent an acquisition orchestrated through emergency arrangement by the Swiss government, aimed at averting its failure.
On 23 March Federal Reserve System, obliged to maintain stable value of the currency, decided to raise the federal rates, resulting in the increase of SOFR from 4.55\% to 4.8\%. This escalation in borrowing costs precipitated another wave of turbulence in the global financial system.
On Friday, 24 March, the market experienced a surge in insurance cost against the collapse of Deutsche Bank, the largest bank in the largest European economy: the price of the 5-year Credit Default Swap for its debt soared by 70\%.
International banking indices were subjects of significant declines simultaneously. On the following trading day, March 27th, financial systems entered a period of chaotic fluctuations, with SOFR oscillating wildly and reaching peaks of 4.84\% and 4.87\% on March 28th and 31st, respectively. The latter spike represents a significant deviation from the plateau, primarily hovering around 4.80-4.81\%. This deviation signals a concerning scenario, highlighting significant challenges encountered by the central bank in preserving banking system stability while adhering to targeted interest rate levels. Such circumstances pose inherent risks, as an excessive increase in interest rates can precipitate a detrimental cycle between the economy and the banking system resulting in a severe crash, as delineated in the framework proposed in 
our previous work  \cite{fortuna_unified_2023}.
\begin{figure}
    \centering
    \includegraphics[width=0.49\textwidth]{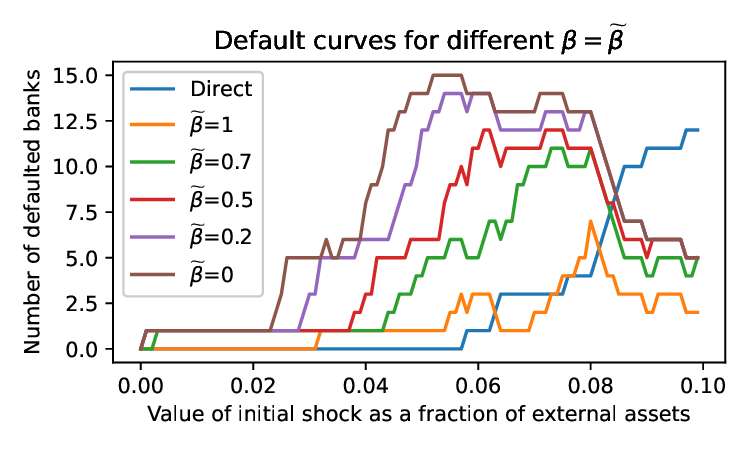}
    \includegraphics[width=0.49\textwidth]{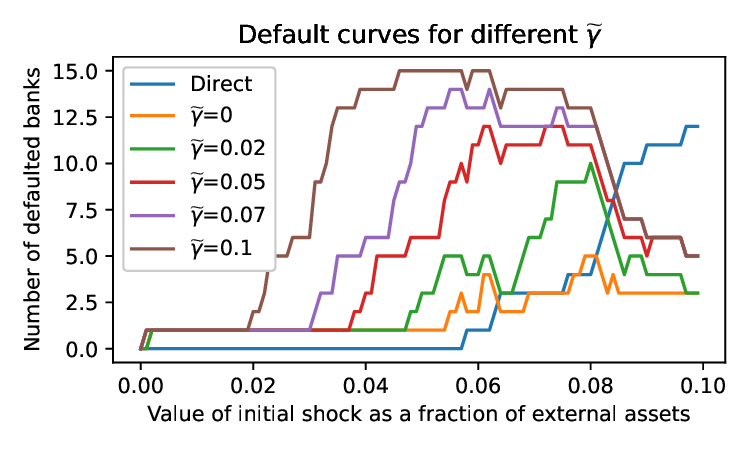}
    \caption{The impact of interest rate feedback on the number of defaults.
    The chart illustrates the relationship between the number of defaulted institutions and varying percentages of shock to external assets, computed using a reduced-form model incorporating interest rate feedback (Eq.~(\ref{eq:ir_feedback})). The blue curve represents direct defaults triggered by the shock, while distinct curves depict indirect defaults, each corresponding to different parameter values. For the left panel, computations were performed with parameter values $(T-t) = 1, \ \gamma_j \equiv 20, \ \widetilde{\gamma} = 0.05,$ and various curves corresponding to different values of the recovery parameter $\beta_j = \widetilde{\beta}$. The quantity of defaults decreases with $\beta_j$ as it determines the assets retrieved in case of default. Similarly, for the right panel, computations were conducted with parameter values $(T-t) = 1, \ \gamma_j=20, \ \beta_j=\widetilde{\beta} = 0.5,$ with various curves corresponding to different values of the scale parameter $\widetilde{\gamma}$. The amount of defaults increases with $\widetilde{\gamma}$ as it describes the pace of valuation decline.}
    \label{fig:rf_ir_feedback}
\end{figure}

The evolution of the American financial system according to the interest rates feedback model (Eq.~(\ref{eq:ir_feedback})) is illustrated in Fig.~\ref{fig:rf_ir_feedback}.
It is notable that the parameter values, particularly $\widetilde{\gamma}$,  are selected judiciously, indicating that the observed effect does not stem from an unreasonable exaggeration in magnitude, but rather arises from the inherent characteristics of the phenomenon under investigation. Analogously to the dynamics of a single institution credit spread $s_t$ in Eq.~(\ref{eq:spreads_limits}), the change in the interest rate $\Delta r_t$ is bounded from above by the value of parameter $\widetilde{\gamma}$, although it is important to mention that the system usually do not reach this extreme. In the examined case, the parameter values ranged between 0-0.1. This range is justified, as for example from the January 2022 to January 2024 the Federal Funds Rate varied from 0.0008 up to over 0.05, while historically reaching even nearly 0.2. 
The integration of interest rates into the model results in a significant alteration in the system's characteristics. In the absence of an interest rate feedback loop, as illustrated in Fig. \ref{fig:rf_params}, discernible yet constrained shifts and amplifications around critical points at 6\% and 8\% indicate a limited systemic impact. However, incorporating interest rate dependency from Eq. (\ref{eq:ir_dynamics}) leads to notable transpositions of these critical points. 
Specifically, the mild critical point transitions from 6\% to a range spanning 2\%-5\%, while the severe crisis point shifts from 8\% to a range of 3\%-7\%, attributable to extensive cascading failures induced by systemic feedback mechanisms.
This difference underscores a fundamental discrepancy between models with and without interest rate dependencies on market conditions. The crunch of the financial system varies significantly, as the inclusion of interest rates precipitates a collapse driven by the common exposure to interest rates and indirectly by the interplay between interest rates and the banking environment. Consequently, neglecting interest rates and their interconnections with the banking system in modeling can obscure risk recognition, resulting in a substantial underestimation of actual risk and eventual system collapse due to decisions predicated on inadequate analysis.

\section{Discussion}
\label{sec:discussion}

In this work, the reduced form network valuation model with interest rate feedback was constructed and applied to evaluate the resilience of the United States' financial system.  The significant role of the mutual funds in the financial system was identified, illustrating the fact that a crucial component in the banking network is external to the set of banks. Furthermore, the substantial impact of debt valuation processes, particularly with recovery and credit quality parameters, was demonstrated, showcasing the adaptability of the proposed approach across diverse modeling environments, in contrast to the linear DebtRank framework.
Finally, the complex relationship between the interest rates and the banking environment was integrated into the model dynamics, uncovering the potential for cascading failures and a collapse of financial system stemming from interdependencies among its participants, a phenomenon not captured by other models.  These results underscore the complexity inherent in the study of systemic risk, where the exclusion of certain factors can lead to significant deviations in risk assessments, posing a danger of misjudging risks.
Furthermore, they lay the foundations for integrating crucial industry processes such as credit quality assessment and stochastic projection of interest rates into systemic risk quantification, through the bond of reduced form models.
Recent events have repeatedly emphasized the importance of this research area, particularly with the emergence of the  systemic risk weaponisation phenomenon \cite{quaglia_weaponisation_2023}.
The examples comes from all across the globe, ranging from the Taiwanese factories constituting a critical node in  semiconductors trade \cite{colon_systemic_2023}, through the Eurasian gas pipelines bringing fire to the Ukrainian fields \cite{zhou_dynamics_2023}, up to the digital banking infrastructure utilised by Western countries  to harm their adversaries \cite{qureshi_russiaukraine_2022}. 
Understanding the dynamics and vulnerabilities of the global banking system  provides the means to inflict damage on a scale disproportionate to the attacker's resources, as evidenced by the actions of the Houthi militias on the Red Sea \cite{elroi_risk_2024}.
Therefore, it is crucial to continue research in this area to preserve the global security,  as threats extend beyond economic ramifications to physical levels through the chain of dependencies constituting a part of, nomen omen, systemic risk.

\section{Acknowledgements}
This work was supported by the Polish Ministry of Science and Higher
Education (MNiSW) core funding for statutory R\&D activities. Access to the Eikon Reuters analytical platform for data collection has been provided by Wroclaw University of Economics and Business. 
K.F was supported by the grant number 50SD/002624~MPK:~9130740000.

\section{Author contributions statement}
Conceptualization: K.F. and J.S.; implementation and visualisation: K.F.; analysis of the results: K.F. and J.S; original draft preparation: K.F.;  supervision, review and editing: J.S. All authors have read and agreed to the published version of the manuscript.

\section{Additional information}
The authors declare no conflict of interest. The founders had no role in the design of the study; in the collection, analyses, or interpretation of data; in the writing of the manuscript; or in the decision to publish the results.

\appendix
\section{Towards Model Calibration}
\label{subsec:Calibration}

Starting from the assets identity equation
\begin{equation}
    A_i (t) = A_{i}^{e} (t) \mathbb{V}_{i}^{e} \left( \mathbf{E} (t)  \right)  + \mathlarger{\sum}_{j=i}^{n} A_{ij} (t) \mathbb{V}_{ij} \left( \mathbf{E} (t) \right),
\end{equation}
the unknown value of external assets is derived:
\begin{equation}
    A_{i}^{e} (t) \mathbb{V}_{i}^{e} \left( \mathbf{E} (t)  \right) = A_i (t) - \mathlarger{\sum}_{j=i}^{n} A_{ij} (t) \mathbb{V}_{ij} \left( \mathbf{E} (t)  \right).
\end{equation}
Furthermore, it is frequently observed, especially in the context of non-clearing contagion algorithms, that asset dynamics are modeled in relation to the initial state $A_{ij}(0)$, thereby assuming
\begin{equation}
\label{eq:initial_conditions}
    \mathbb{V}_{i}^{e} (E_{j}(0)) = \mathbb{V}_{ij}(E_{j}(0)) = 1.
\end{equation}
This holds true for linear and various versions of DebtRank \cite{battiston_debtrank_2012, bardoscia_correction_2015, bardoscia_distress_2016}, as well as Furfine \cite{furfine_interbank_1999} methods. Such an approach proves to be very natural and reasonable, especially in the context of stress test-focused research. The initial state, derived from market data, is presumed to be at the point of equilibrium. Subsequently, a shock is applied, and its propagation through financial dependencies, along with its ultimate impact on the system, is systematically assessed.
This methodological framework will be employed in  our work. However, it is important to note that the proposed solution is not exhaustive, and alternative approaches can be considered as well.

For linear DebtRank (\ref{eq:linear_dr}), the condition is given straightforward, as

\begin{equation}
    \mathbb{V}_{ij}(E_j (0)) = \min \left[ \frac{E_j^+ (0)}{E_j (0)}, 1 \right] = \min \left[ 1, 1 \right] = 1.
\end{equation}
In case of recovery DebtRank~(\ref{eq:rec_dr_noncal}), it must hold that

\begin{align}
    1 = \mathbb{V}_{ij}(E_j (0)) &= \min \left[ \alpha_j \frac{E_j^+ (0)}{E_j (0)} + \beta_j \frac{A_j^+ (0)}{A_j (0)}, 1 \right] \nonumber \\
    &= \min \left[ \alpha_j + \beta_j, 1 \right].
\end{align}
In order to satisfy the condition above, the recovery parameter is set to $\beta_j = 1 - \alpha_j.$

In case of valuation functions  defined by the reduced form models as in Eq.~(\ref{eq:rf_valuation_functions}), the conditions in Eq.~(\ref{eq:initial_conditions})
can be expressed as 
\begin{equation}
    \Delta r_0 = s_0 = 0.
\end{equation}
Inserting this into the definition of risk spread $s_t$ and interest rates $\Delta r_t$ given in Eq ~(\ref{eq:rf_dr_noncal}) yields
\begin{widetext}   
\begin{align}
    0 &= s_0 =  \max \left[ \gamma_j  \left(1 -  \alpha_j \frac{ \min \left[ E_j^+ (0), \ E_j (0) \right]}{E_j (0)} \right)
     \left(1 - \beta_j \frac{\min \left[ A_j^+ (0), \ A_j (0)/ \beta_j \right]}{A_j (0)} \right), 0 \right] \nonumber \\
    0 &= \Delta r_0 =  \max  \left[   \widetilde{\gamma} \left(1 - 
    \widetilde{\alpha} \frac{ \min \left\{ \norm{ \mathbf{E}^+ (0) }, \norm{ \mathbf{E}(0) }  \right\} }
    { \norm{ \mathbf{E}(0) }} \right) \left(1 - \widetilde{\beta} \frac{ \min \left\{ \norm{ \mathbf{A}^+ (0) }, \widetilde{\beta}^{-1} \norm{ \mathbf{A}(0) }  \right\}}{ \norm{ \mathbf{A}(0) }} \right)  , 0 \right].
\end{align}
\end{widetext}
The condition for $s_0$ can be reduced to
\begin{equation}
    0  = \max \left[   - \gamma_j \left(1 - \alpha_j  \right) \left(1 - \beta_j  \right),  0 \right].
\end{equation}
In order to meet the above criterion, the parameter $\alpha_j$ is fixed to $\alpha_j=1$. The rationale behind the proposed approach is presented below.
\color{black}

In the terminology of credit valuation embedded in the Duffie-Singleton model (\ref{eq:duffie_singleton}), the approach proposed in Eq.~(\ref{eq:rf_dr_noncal}) for the risk spread $s(t) = \lambda(t)(1 - \delta_t)$ can be expressed as follows:
\begin{equation}
    \lambda(t)(1 - \delta_t) = \gamma_j \left(1 - \alpha_j \frac{E_j^+(t)}{E_j(0)}\right) \left(1 - \beta_j \frac{A_j^+(t)}{A_j(0)}\right).
\end{equation}
Here, the recovery rate $\delta_t$ is modeled dynamically by the term $\beta_j \frac{A_j^+(t)}{A_j(0)}$, while the hazard rate component $\lambda(t)$ is delineated by $\gamma_j \left(1 - \alpha_j \frac{E_j^+(t)}{E_j(0)}\right)$.
The parameter $\alpha_j$, constrained within the interval $[0, 1]$, presents limited modeling prospects. It reflects the dependency of a company's default probability on its equity. However, as described in detail in Subsec.~\ref{subsec:rf_gammabeta}, the essential characteristics of default probability are predominantly governed by the $\gamma_j$ parameter, which offers greater adaptability across diverse real-world scenarios compared to $\alpha_j$. 
Particularly, $\gamma_j$ is responsible for the rate of decline in valuation of company debt instruments, and provides full flexibility of valuation boundary up to 0 in the limit, a feature that cannot be achieved by the $\alpha_j$ (a thorough mathematical analysis is conducted in Subsec.~\ref{subsec:rf_dr_met}).
Consequently, it is sensible to calibrate the model with the use of the $\alpha_j$ parameter. 
This approach maintains the model's flexibility and comprehensive reflection of all pertinent aspects of the Duffie-Singleton framework, while also addressing the requirements of our research. Indeed, under the proposed parameterization, strength of the recovery effect is governed by
$\beta_j \frac{A_j^+(t)}{A_j(0)}$, while  the intensity of the hazard rate is regulated by $\gamma_j.$ This delineation of responsibilities among model parameters aligns with real-world dynamics and facilitates clear interpretation. Alternative solutions that introduce interdependencies among parameters risk compromising this clarity and coherence.

The presented  reasoning can be analogously  extended to the case of the interest rate $\Delta r_0$, ultimately resulting  in an aggregated condition  
$\widetilde{\alpha} = \alpha_j = 1.$

\end{document}